\documentclass[reprint,preprintnumbers,showpacs,nofootinbib,amsmath,amssymb,aps,prd,letterpaper,
]{revtex4-1}
\usepackage{graphicx}
\usepackage{dcolumn}
\usepackage{bm}
\usepackage{subfigure}
\usepackage{color}
\usepackage[active]{srcltx}
\usepackage{amsmath,amsfonts,amssymb,amsthm,amstext,amscd,eucal,srcltx}
\usepackage{epsfig,graphicx,bm}
\usepackage{epstopdf, epsf}
\usepackage{dcolumn}
\usepackage{hyperref}
\everymath{\displaystyle}
\begin{document}
\title{A new phase of scalar field with a kinetic term non-minimally coupled to gravity}
\author{Amir Ghalee}
\affiliation{{Department of Physics, Tafresh  University,
P. O. Box 39518-79611, Tafresh, Iran}}
\begin{abstract}
We consider the dynamics of a scalar field non-minimally coupled to gravity in the context of cosmology.
It is demonstrated that there exists a new phase for the scalar field, in addition to the inflationary and dust-like (reheating period) phases. Analytic expressions for the scalar field and the Hubble parameter, which describe
the new phase are given. The Hubble parameter indicates an accelerating expanding Universe. We explicitly show that the scalar field oscillates with time-dependent frequency. Moreover, an interaction between the scalar field in the new phase and other fields is discussed.
It turns out that the parametric resonance is absent, which is another crucial difference between the dynamics of
the scalar field in the new phase and dust-like phase.
\end{abstract}
\maketitle
\section{\label{sec:level1}INTRODUCTION}
To solve the flatness problem and the horizon problem in cosmology, Alan Guth introduced the inflation paradigm
\cite{guth}. The simple model, which is a single scalar field (inflaton) with minimally
coupling to gravity, is described by an action
\begin{equation}\label{standard-action}
S=\int d^{4}x\sqrt{-g}\left[-\frac{R}{2\kappa^{2}}-\frac{1}{2}g^{\mu\nu}\partial_{\mu}\varphi\partial_{\nu}\varphi-V(\varphi)
\right],
\end{equation}
where $\kappa^{2}\equiv8\pi G$. Although model \eqref{standard-action} with quadratic potential, $V(\varphi)=\frac{1}{2} m^{2}\varphi^{2}$, is consistent with the cosmic microwave background \cite{CMB-data}, many models have been proposed to produce the inflation era \cite{bassett}. Regarding the motivations, many attempts are devoted to derive the inflation era by more fundamental principles or find connections between inflaton and other fields
that are used in particle physics(e.g. Higgs field) \cite{bassett,mukhanov,Germani}.\\
An intuitive picture of the dynamics of inflaton is as follows:
the inflaton field rolls slowly down its potential (inflationary period),
eventually the field(s)oscillates around the minimum of its potential and decays into light particles (reheating period).
It has been assumed that the reheating period takes place just after the inflationary period. Also, since
the inflaton has lost its energy, it behaves like dust matter in the reheating period.\\
For the quadratic potential, this
intuitive picture is supported by analytic methods \cite{mukhanov}, ( see \cite{quartic} for the quartic potential).
Usually model builders use the stated scenario for their models ( see \cite{ghalee} for a different scenario). Then, they set limits on the parameters of the models.
So, one can divide the dynamics of inflaton in two phases; inflationary phase and matter dominate
phase(near minimum of the potential).\\
In this work, we show that in the following effective action
\begin{equation}\label{action}
S=\int d^{4}x\sqrt{-g}\left[-\frac{R}{2\kappa^{2}}-\frac{1}{2}(g^{\mu\nu}-\alpha^{2}G^{\mu\nu})\partial_{\mu}\varphi\partial_{\nu}\varphi-V(\varphi)
\right],
\end{equation}
where $G^{\mu\nu}$ is Einstein's tensor, the scalar field has a new phase. Germani et.al \cite{Germani}, proposed the above action with quartic potential, $\lambda\varphi^{4}$, in the context of
the new Higgs inflation. The cosmological perturbation of the model and reheating period of the model have been studied in \cite{Germani2} and
\cite{mohseni} respectively.\\
This paper is organized as follows: in \S II we briefly review the model and obtain equations, then we qualitatively discuss why
the scalar field of model \eqref{action}, has a new phase for any typical potential,$V(\varphi)$. In \S III analytic solutions for
the Hubble parameter and the scalar field are provided for the quadratic potential. The solutions describe the new phase of the scalar field. In \S \textrm{IV}
we investigate an implication of the new phase by considering an interaction between a relativistic field and the scalar field. To introduce reader to the method that is used in this paper, we re-derive
solution of the scalar field of the model \eqref{standard-action} in reheating period in Appendix A, which is identical
with the solution in the textbook \cite{mukhanov}. Appendix B is devoted to some properties of the Fresnel integrals.
\section{\label{sec:level1}	Evidences for existence of the new phase}
To obtain field equation and the Friedman equation of the model in FRW background metric,
we use the ADM formalism with the following metric ansatz \cite{wald}
\begin{equation}\label{metric}%
ds^{2}=-N(t)dt^{2}+a(t)^{2}\delta_{ij}dx^{i}dx^{j}.
\end{equation}
By inserting this expression into \eqref{action} we have \cite{Germani,Germani2}
\begin{equation}\label{ADM action}%
S=\int dta^{3}\left[-3\frac{H^{2}}{\kappa^{2}N}+\frac{1}{2}\frac{\dot{\varphi}^{2}}{N}+\frac{3}{2}\frac{H^{2}\alpha^{2}\dot{\varphi}^{2}}{N^{3}}-NV(\varphi)
\right],
\end{equation}
by varying the action\eqref{ADM action} with respect to the laps $N$ and $\varphi$ and setting $N$ to 1, we obtain
\begin{equation}\label{Friedmann1}
\begin{split}
&H^{2}=\frac{\kappa^{2}}{6}\left[\dot{\varphi}^{2}(1+9\alpha^{2}H^{2})+2V(\varphi)\right] ,\cr
&\ddot{\varphi}(1+3H^{2}\alpha^{2})+3H\dot{\varphi}(1+3H^{2}\alpha^{2})+6\dot{\varphi}H\dot{H}\alpha^{2}=-\frac{dV(\varphi)}{d\varphi} .
\end{split}
\end{equation}
The behavior of the scalar field can be divided into two regimes \cite{note1}
\begin{itemize}
  \item If $\alpha H\ll1$, from \eqref{Friedmann1} we have
\begin{subequations}
\begin{align}
        &H^{2}=\frac{\kappa^{2}}{6}\left[\dot{\varphi}^{2}+2V(\varphi)\right]\label{1var1},\\
        &\ddot{\varphi}+3H\dot{\varphi}=-\frac{dV(\varphi)}{d\varphi} \label{1var2}.
\end{align}
\end{subequations}
\end{itemize}
Equations \eqref{1var1} and \eqref{1var2} are the same as equations which are derived from \eqref{standard-action} and studied in textbooks \cite{mukhanov,Weinberg}. Since we
want to compare other regime with this regime, we quote the main results.\\
The second term on the left hand side of \eqref{1var2}, which is \textit{always} positive, acts as a dissipative force. Therefore, the scalar filed rolls toward
the minimum of the potential, and it behaves as a dust matter. For the simplest potential,$V(\varphi)=\frac{1}{2}m^{2}\varphi^{2}$, it has been shown that \cite{mukhanov}
\begin{equation}\label{1field-approx1}
H(t)=\frac{2}{3t}\quad,\quad \varphi(t)=\frac{\sqrt{6}H(t)}{\kappa m}\cos\left[mt+g(0)\right]
\end{equation}
where $g(0)$ is an arbitrarily constant. It is important to note that expressions in \eqref{1field-approx1} are only valid for $H(t)<m$, so
higher order terms, have been neglected \cite{mukhanov}.
   \begin{itemize}
\item If $\alpha H\gg1$, from \eqref{Friedmann1} we have
\begin{subequations}
\begin{align}
        &H^{2}=\frac{\kappa^{2}}{6}\left[\dot{\varphi}^{2}9\alpha^{2}H^{2}+2V(\varphi)\right]\label{2var1},\\
       &\ddot{\varphi}-3H\dot{\varphi}w_{eff}=-\frac{1}{3H^{2}\alpha^{2}}\frac{dV(\varphi)}{d\varphi}\label{2var2},
\end{align}
\end{subequations}
where $w_{eff}$ is the effective equation of state
\begin{equation}\label{variable-def}
w_{eff}=-1-\frac{2}{3}\frac{\dot{H}}{H^{2}}\quad.
\end{equation}
   \end{itemize}
The inflationary phase of this regime, $w_{eff}\rightarrow -1$, was studied in \cite{Germani,Germani2}. Note that during the inflationary phase, the second term on the left hand side in \eqref{2var2} is positive,
and acts as a dissipative force. So, again, the scalar filed rolls toward the minimum of the potential , but at $H=2/3t$, it is vanished and, after some time,
its sign may be changed and act as a driving force. Another point is that the
right hand side of \eqref{2var2}, depends on $H$, which is not constant in this period.
This behavior, shows that we need a different analysis for this regime.

\section{\label{sec:level1}Explicit solutions for the quadratic potential}
In this section we consider the simple potential $V(\varphi)=\frac{1}{2} m^{2} \varphi^{2}$, then we seek solutions for the equations. We use
the so-called averaging method \cite{Guckenheimer}, which is used in study of nonlinear differential equations. In Appendix A, we use this method to re-derive \eqref{1field-approx1}. \\
Let us define a new time variable $\tau$ as
\begin{equation}\label{time-def}
t=\int\frac{3\alpha H}{m}d\tau,
\end{equation}
It is worth to mention that this step is only a trick to solve the equations,
when we obtain the solutions we'll back to the original time variable, $t$.\\
Combining Eqs. \eqref{2var1},\eqref{2var2} and \eqref{time-def}, with the quadratic potential, gives
\begin{equation}\label{Friedmann3}
\begin{split}
&H^{2}=\frac{\kappa^{2}}{6}m^{2}\left(\varphi'^{2}+\varphi^{2}\right) ,\cr
&\varphi''+\left(\frac{H'}{H}+\frac{9H^{2}\alpha}{m}\right)\varphi'=-3\varphi,
\end{split}
\end{equation}
where prime denotes the differential with respect to $\tau$.\\
In order to use the method of averaging, we define the following relations \cite{Guckenheimer}
\begin{subequations}
\begin{align}
        &\varphi=\frac{\sqrt{6}H}{\kappa m}\cos\left(\tau+f(\tau)\right)\label{var1},\\
        &\varphi'=\frac{-\sqrt{6}H}{\kappa m}\sin\left(\tau+f(\tau)\right)\label{var2},
\end{align}
\end{subequations}
where $f(\tau)$ is an arbitrary function. Differentiation of the righthand side
of \eqref{var1} must be equals to the righthand side of \eqref{var2}. Hence we have
\begin{equation}\label{con1}
H'\cos\left(\tau+f(\tau)\right) =Hf'\sin\left(\tau+f(\tau)\right) ,
\end{equation}
Substitution of \eqref{var1} and \eqref{var2} into \eqref{Friedmann3} results in
 \begin{equation}\label{con2}
\begin{split}
&2H'\sin\left(\tau+f(\tau)\right)+Hf'\cos\left(\tau+f(\tau)\right)\cr
&+\frac{9H^{3}\alpha}{m}\sin\left(\tau+f(\tau)\right)-2H\cos\left(\tau+f(\tau)\right)=0 .
\end{split}
\end{equation}
$H'$ and $f'$ can be found from Equations \eqref{con1} and \eqref{con1} as
\begin{equation}\label{algebra}
\begin{split}
&f'=\frac{\cos\left(\tau+f(\tau)\right)}{1+\sin^{2}\left(\tau+f(\tau)\right)}\left[2\cos\left(\tau+f(\tau)\right)-\frac{9H^{2}\alpha}{m}\sin\left(\tau+f(\tau)\right)\right]\cr
\\
&H'=\frac{H\sin\left(\tau+f(\tau)\right)}{1+\sin^{2}\left(\tau+f(\tau)\right)}\left[2\cos\left(\tau+f(\tau)\right)-\frac{9H^{2}\alpha}{m}\sin\left(\tau+f(\tau)\right)\right] .
\end{split}
\end{equation}
So far we have not used any approximation for $H$, and $f(\tau)$. In the method of averaging, equations in \eqref{algebra} are replaced by averaged expressions.
For this goal, note that if we keep $H$ and $f$ fixed \cite{noteref}, the righthand side of \eqref{algebra} are $\pi$-periodic in $\tau$, so we can average over $\tau$.\\
By applying the stated procedure, one can show that
\begin{equation}\label{sin}
\begin{split}
&\left<\frac{\cos^{2}\left(\tau+f(\tau)\right)}{1+\sin^{2}\left(\tau+f(\tau)\right)}\right>=\sqrt{2}-1, \left<\frac{\sin^{2}\left(\tau+f(\tau)\right)}{1+\sin^{2}\left(\tau+f(\tau)\right)}\right>=1-\frac{\sqrt{2}}{2},\cr
&\left<\frac{\cos\left(\tau+f(\tau)\right)\sin\left(\tau+f(\tau)\right)}{1+\sin^{2}\left(\tau+f(\tau)\right)}\right>=0,
\end{split}
\end{equation}
where $< \cdots>\equiv \oint (\cdots)d\tau/\pi$ denotes an average over $\tau$ (but $H$ and $f$ are fixed).
So, from Eqs. \eqref{algebra} and \eqref{sin}, we have
\begin{equation}\label{averagedeqs}
\begin{split}
&f'=2(\sqrt{2}-1)\cr
&H'=-\frac{9H^{3}\alpha}{m}(1-\frac{\sqrt{2}}{2}) .
\end{split}
\end{equation}
The averaged equations are very easy to solve:
\begin{equation}\label{field-approx1}
H(\tau)=\left[\frac{18\alpha}{m}(1-\frac{\sqrt{2}}{2})\tau\right]^{\frac{-1}{2}},\hspace{.1in} f(\tau)=2(\sqrt{2}-1)\tau.
\end{equation}
We can use formula \eqref{time-def} and \eqref{field-approx1} to obtain explicit relation between $\tau$ and $t$ as
\begin{equation}\label{explisit}
t=2\left[\frac{2m}{\alpha}(1-\frac{\sqrt{2}}{2})\right]^{\frac{-1}{2}}\tau^{\frac{1}{2}}.
\end{equation}
By using Eqs. \eqref{field-approx1} and \eqref{explisit} we have
\begin{equation}\label{constraine}
H(t)=\frac{2}{3(2-\sqrt{2})t}
\end{equation}
so, the effective equation of state becomes
\begin{equation}\label{eff2}
w_{eff}=-1-\frac{2}{3}\frac{\dot{H}}{H^{2}}\approx-0.41.
\end{equation}
Moreover, from \eqref{var1} and \eqref{explisit}, we also obtain
 \begin{equation}\label{field2}
\varphi(t)=\frac{\sqrt{6}H(t)}{\kappa m}\cos\left[\frac{m}{2\alpha}(2-\sqrt{2})(\sqrt{2}-\frac{1}{2})t^{2}\right]
\end{equation}
where $H(t)$ is given by \eqref{constraine}.\\
The new phase of the scalar field is described by \eqref{constraine} and \eqref{field2}.
According to \eqref{eff2}, for this new phase we have $w_{eff}<-1/3$, so, the Universe is driven to accelerated expansion phase by
the scalar field. Expression
\eqref{field2}, shows that the scalar field oscillates with time-dependent "frequency", as indicated in Fig.1.  Both of the properties are
more different than the dust-like phase, which is described by \eqref{1field-approx1}.
\begin{figure}[h]
\includegraphics{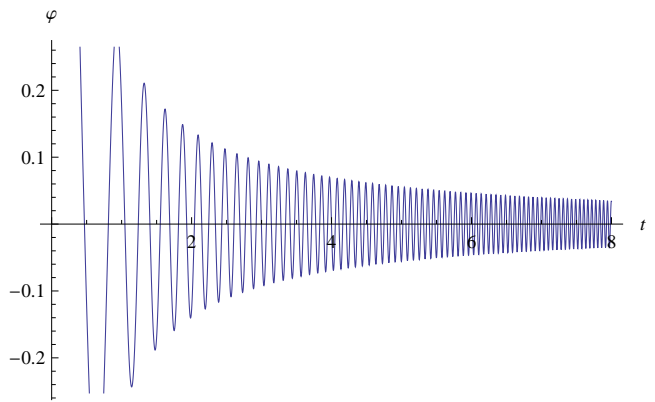}
\caption{\label{fig:epsart}  The scalar field versus $t$ for $\kappa m=10$, $m=10\alpha$ (with $\hbar=c=1$). Compare with Fig[1] in the first and
 the second papers of \cite{mohseni}}
\end{figure}
\section{\label{sec:level1}An Implication}
In this section, to show an implication of properties of the new phase, we will consider an interaction between a matter field,$\chi$, and the scalar
field.\\
The decay of the scalar field into a relativistic field can be described by \cite{mukhanov}
\begin{equation}\label{lag}
\mathcal{L}_{int}=-\frac{1}{2}g^{2}\varphi^{2}\chi^{2}.
\end{equation}
It has been shown that if the interaction takes place during the dust-like phase, the parametric resonance instability is occurred \cite{parametric resonance,mukhanov}.\\
Notice that this result is obtained if the
expansion of the universe is neglected. The stated assumption, seems to be a reasonable condition, at least for the first approximation, if the rate of interaction
is too fast compared to the Hubble expansion time, see discussions about this note in \cite{mukhanov}. As mentioned in \cite{mukhanov}, what one actually
finds from present of the parametric resonance, is that the perturbative analysis is rather misleading, during dust-like phase of the inflaton.\\
Here,our aim is to quest for the parametric resonance, during the new phase of the scalar field.\\
If the other scalar field, $\chi$, is decomposed into Fourier modes as
\begin{equation}\label{Fou}
    \chi(\vec{X},t)=\frac{1}{(2\pi)^{3/2}}\int d^{3}k\left(\chi^{*}_{k}(t)e^{i\vec{k}.\vec{X}}+\chi_{k}(t)e^{-i\vec{k}.\vec{X}}\right),
\end{equation}
then using \eqref{field2} and \eqref{lag}, the following equation for the Fourier modes is obtained
\begin{equation}\label{re1}
\ddot{\chi_{k}}+\left(k^{2}+g^{2}A^{2}\cos^{2}(M^{2}t^{2})\right)\chi_{k}=0,
\end{equation}
where
\begin{equation}\label{re2}
A^{2}\equiv\frac{6}{\kappa^{2}}\frac{H^{2}}{m^{2}}, \qquad M^{2}\equiv\frac{m}{2\alpha}(2-\sqrt{2})(\sqrt{2}-\frac{1}{2}).
\end{equation}
Now, following \cite{mukhanov}, we will neglect the expansion of space. So, If we define the following variables
\begin{equation}\label{re3}
\xi\equiv\sqrt{2}Mt, \quad \omega^{2}\equiv\frac{k^{2}+g^{2}A^{2}}{2M^{2}},\quad \varepsilon\equiv\frac{g^{2}A^{2}}{4M^{2}},
\end{equation}
it thus follows that
\begin{equation}\label{re4}
\frac{d^{2}\chi_{k}}{d\xi^{2}}+(\omega^{2}+\varepsilon\cos\xi^{2})\chi_{k}=0.
\end{equation}
To obtain an approximate solution for $\chi_{k}$, we expand $\chi_{k}$ as
\begin{equation}\label{re5}
 \chi_{k}=\chi^{0}_{k}+\varepsilon\chi^{1}_{k}+\ldots\quad.
\end{equation}
The expression \eqref{re5} is valid, if $\chi^{1}_{k}$ has not terms that grow without bound as $t\rightarrow\infty$. So, it is necessary to cross-check this condition
after we obtain an explicit expression for $\chi^{1}_{k}$.\\
Substituting \eqref{re5} into \eqref{re4}, and keeping all terms up to second-order in $\varepsilon$, yields
\begin{equation}\label{re6}
    \begin{split}
      &\frac{d^{2}\chi^{0}_{k}}{d\xi^{2}}+\omega^{2}\chi^{0}_{k}=0 \\
      &\frac{d^{2}\chi^{1}_{k}}{d\xi^{2}}+\omega^{2}\chi^{1}_{k}=-\chi^{0}_{k}\cos\xi^{2}.
\end{split}
\end{equation}
The first equation in \eqref{re6} gives
\begin{equation}\label{re7}
    \chi^{0}_{k}=b_{1}\sin\omega\xi+b_{2}\cos\omega\xi,
\end{equation}
where $b_{1}$ and $b_{2}$ are the integration constants.\\
Using \eqref{re7} and the following variables
\begin{equation}\label{re8}
    u\equiv\sqrt{\frac{2}{\pi}}(\xi+\omega),\qquad v\equiv\sqrt{\frac{2}{\pi}}(\xi-\omega),
\end{equation}
the second equation in \eqref{re6} can be solved as
\begin{equation}\label{re9}
    \begin{split}
\chi^{1}_{k}=&Q_{1}\cos(\omega \xi)+Q_{2}\sin(\omega \xi) -\frac{1}{4\omega}\sqrt{\frac{\pi}{2}}\bigg\{ \\
&\cos(\omega\xi+\omega^{2})[b_{_{1}}C(u)-b_{2}S(u)]+\\
&\sin(\omega\xi+\omega^{2})[b_{_{2}}C(u)+b_{1}S(u)]+\\
&\cos(\omega\xi-\omega^{2})[b_{_{1}}C(v)+b_{2}S(v)]+\\
&\sin(\omega\xi-\omega^{2})[b_{_{2}}C(v)-b_{1}S(v)]+\\
&2C(\frac{u+v}{2})[b_{2}\sin(\omega \xi)-b_{1}\cos(\omega \xi)]\bigg\}.
    \end{split}
\end{equation}
Where $Q_{1}$, $Q_{2}$, are the integration constants, and $C(x)$, $S(x)$ are the Fresnel integrals ( see Appendix B).\\
From properties of the Fresnel integrals\cite{Abramowitz}, we conclude that expression \eqref{re9} has not
any terms that grow without bound as $t\rightarrow\infty$. So, the parametric resonance is absent in \eqref{re1}. Therefore, we expect that the perturbation approximation for $\chi$,
which is given by \eqref{re5}, is assured for all times.\\
Figure 2 compares the approximate solution for $\chi$, to the numerical solution. The two curves are almost indistinguishable.\\
If we used \eqref{re1} instead of \eqref{1var2}, we would have terms that grow without bound as $t\rightarrow\infty$, which
are sources of the parametric resonance during dust-like phase of the scalar field.
 \begin{figure}[t]
\subfigure[][]{ \label{fig:ex2-a}\includegraphics{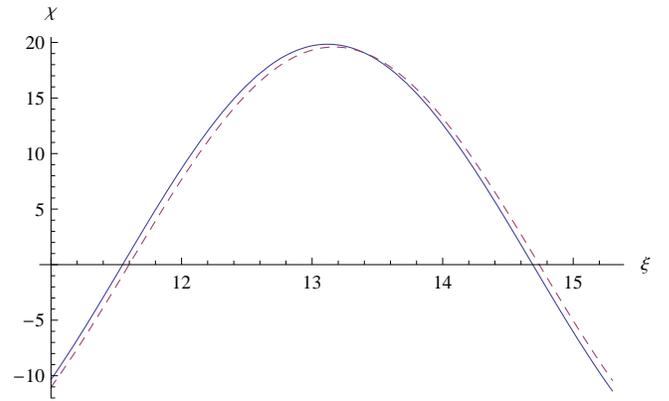}}
\subfigure[][]{\label{fig:ex2-b}\includegraphics{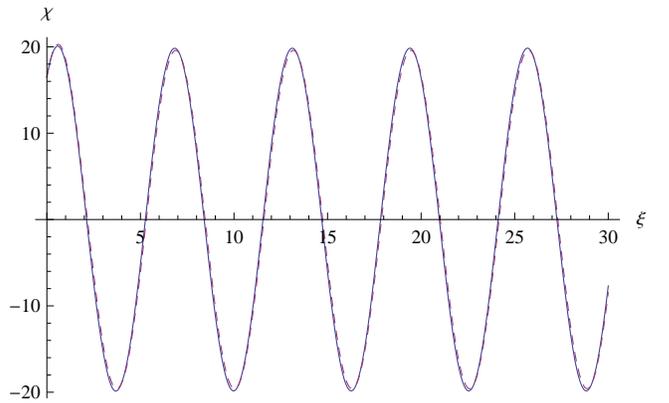}}
\caption{The relativistic field,$\chi$, versus $\xi$. Solid line
represents the numerical solution. Dashed line represents the approximate solution. Where the parameters are chosen such that $\chi=16.5$, $\frac{d\chi}{d\xi} =12$. The two curves are so similar
that it is almost impossible to distinguish between them, as shown in \subref{fig:ex2-b}.}
\label{fig:ex2}
\end{figure}
\begin{acknowledgments}
I would like to thank S. Vasheghani Farahani for read the manuscript. I am grateful for helpful discussions with H.~Mohseni Sadjadi, P.~Goodarzi.
\end{acknowledgments}
\appendix
\section{}
In this appendix, to show the power of the method that is used in this paper, we re-derive \eqref{1field-approx1}, by the averaging method.\\
By introducing $\beta\equiv mt$, and $V(\varphi)=\frac{1}{2}m^{2}\varphi^{2}$, equations \eqref{1var1} and \eqref{1var2} can be written to give
\begin{equation}\label{AFriedmann2}
\begin{split}
&H^{2}=\frac{m^{2}\kappa^{2}}{6}\left[\dot{\varphi}^{2}+\varphi^{2}\right] ,\cr
&\ddot{\varphi}+3\frac{H}{m}\dot{\varphi}=-\varphi\ .
\end{split}
\end{equation}
Now, consider the following relations
\begin{subequations}
\begin{align}
        &\varphi=\frac{\sqrt{6}H}{\kappa m}\cos\left(\beta+g(\beta)\right)\label{Avar1},\\
        &\varphi'=\frac{-\sqrt{6}H}{\kappa m}\sin\left(\beta+g(\beta)\right)\label{Avar2},
\end{align}
\end{subequations}
where $g(\beta)$ is an arbitrary function. Differentiation of the righthand side
of \eqref{Avar1} must be equals to the righthand side of \eqref{Avar2}. Hence we have
\begin{equation}\label{Acon1}
H'\cos\left(\beta+g(\beta)\right) =Hg'\sin\left(\beta+g(\beta)\right) ,
\end{equation}
Substitution of \eqref{Avar1} and \eqref{Avar2} into \eqref{AFriedmann2} yield
 \begin{equation}\label{Acon2}
\begin{split}
&H'\sin\left(\beta+g(\beta)\right)+Hg'\cos\left(\beta+g(\beta)\right)\cr
&+\frac{3H^{2}}{m}\sin\left(\beta+g(\beta)\right)=0.
\end{split}
\end{equation}
By algebraic manipulations, $H'$ and $g'$ can be found from Equations \eqref{Acon1} and \eqref{Acon2} as
\begin{equation}\label{Aalgebra}
\begin{split}
&g'=-\frac{3H}{2m}\sin\left(\beta+g(\beta)\right)\cos\left(\beta+g(\beta)\right)\cr
\\
&H'=-\frac{3H^{2}}{m}\sin^{2}\left(\beta+g(\beta)\right) .
\end{split}
\end{equation}
These expression for $H$, and $g(\beta)$ are exact.\\
The advantage of the averaging method is that an approximation to the solution of \eqref{Aalgebra} can be obtained by
replacing \eqref{Aalgebra} with its averaged equations as follows:
if we keeping $H$ and $g$ fixed, the righthand side of \eqref{Aalgebra} are $\pi$-periodic in $\beta$. By noting that
\begin{equation}\label{Asin}
\left<\sin^{2}\left(\beta+g(\beta)\right)\right>=\frac{1}{2},\quad
\left<\cos\left(\beta+g(\beta)\right)\sin\left(\beta+g(\beta)\right)\right>=0,
\end{equation}
where $< >$ is used to indicate that we average over $\beta$, the averaged equations can be solved as
\begin{equation}\label{Afield-approx1}
H(t)=\frac{2}{3t}\quad,\quad g(t)=g(0) ,
\end{equation}
where g(0) is an arbitrary constant. Using \eqref{Avar1} and \eqref{Afield-approx1}, we have
\begin{equation}\label{Afield2}
\varphi(t)=\frac{\sqrt{6}H(t)}{\kappa m}\cos\left[mt+g(0)\right]
\end{equation}
where $H(t)$ is given by \eqref{Afield-approx1}.\\
According to the averaging method, the above expressions for Hubble parameter and the scalar field are valid
for $H(t)<m$. The results are the same as those obtained in \cite{mukhanov}, which are also valid for $H(t)<m$.
 \section{}
 The Fresnel integrals are defined by \cite{Abramowitz}
\begin{equation}\label{fre}
\begin{split}
&C(x)\equiv\int^{x}_{0}\cos(\frac{1}{2}\pi x^{2})dx \\
&S(x)\equiv\int^{x}_{0}\sin(\frac{1}{2}\pi x^{2})dx
\end{split}
\end{equation}
A series expansion for $x<1$ gives
\begin{equation}\label{serfre}
\begin{split}
&C(x)=\sum^{\infty}_{n=0}\frac{(-1)^{n}(\pi/2)^{2n}}{(2n)!(4n+1)}\hspace{1mm}x^{4n+1}, \\
&S(x)=\sum^{\infty}_{n=0}\frac{(-1)^{n}(\pi/2)^{2n+1}}{(2n+1)!(4n+3)}x^{4n+3}.
\end{split}
\end{equation}
Therefore, $\lim_{x\rightarrow0}C(x)=\lim_{x\rightarrow0}S(x)=0$.\\
Asymptotic expansion of the integral are given by
\begin{equation}\label{asyfre}
\begin{split}
&C(x)=\frac{1}{2}+\frac{1}{\pi x}\sin(\frac{1}{2}\pi x^{2}),\\
&S(x)=\frac{1}{2}-\frac{1}{\pi x}\cos(\frac{1}{2}\pi x^{2}).
\end{split}
\end{equation}
Hence, $\lim_{x\rightarrow\infty}C(x)=\lim_{x\rightarrow\infty}S(x)=1/2$.
\bibliography{apssamp}

\begin{thebibliography}{99}
\bibitem{note1}
It is worth noting that, in this work, we focus on solving the equations and
study some physical consequences of the solutions. For this purpose, the constrains on the parameters of the model, e.g $\alpha$, which
was obtained in \cite{Germani2}, are not important for us. So, our classification of the regimes, is not in chronological order. As we will see, this
classification is suitable to discuss the difference between\eqref{standard-action} and \eqref{action}.If one like to stick to the constrains on  parameters, which are
obtained in Ref. \cite{ Germani,Germani2,mohseni},we have  the  new phase,
\eqref{1field-approx1}, and then dust-like phase, \eqref{field2}.
\bibitem{noteref}
Physicaly, it means that we integrate out "fast" modes, so, in fact, $H$ and $f(\tau)$ are slowly varying functions. For further information
about this subject see \cite{Guckenheimer}.
\end{thebibliography}

\end{document}